\documentclass{ws-procs9x6}

\def\trimmarks{}
\def\draftnote{\hfill}

\begin{document}

\title{Neutrinos in Physics and Astrophysics}

\author{GEORG~G.~RAFFELT}

\address{Max-Planck-Institut f\"ur Physik
(Werner-Heisenberg-Institut)\\
F\"ohringer Ring 6, 80805 M\"unchen, Germany\\
E-mail: raffelt@mppmu.mpg.de}

\maketitle

\vskip-12pt
\vbox to 0pt{{\ }\newline\vskip-7.5cm \noindent
\normalsize\rm Contribution to the Proceedings of 
{\it Texas in Tuscany:
XXI Symposium on Relativistic Astrophysics}, 
9--13 December 2002,
Florence, Italy.\vfil}

\abstracts{The observed flavor oscillations of solar and atmospheric
neutrinos determine several elements of the leptonic mixing matrix,
but leave open the small mixing angle $\Theta_{13}$, a possible
CP-violating phase, the mass ordering, the absolute mass scale
$m_\nu$, and the Dirac vs.\ Majorana property.  Progress will be made
by long-baseline, tritium endpoint, and $2\beta$ decay
experiments. The best constraint on $m_\nu$ obtains from cosmological
precision observables, implying that neutrinos contribute very little
to the dark matter.  However, massive Majorana neutrinos may well be
responsible for ordinary matter by virtue of the leptogenesis
mechanism for creating the baryon asymmetry of the universe.  In
future, neutrinos could play an important role as astrophysical
messengers if point sources are discovered in high-energy neutrino
telescopes. In the low-energy range, a high-statistics observation of
a galactic supernova would allow one to observe directly the dynamics
of stellar collapse and perhaps to discriminate between certain mixing
scenarios. An observation of the relic neutrinos from all past
supernovae has come within reach.}

\section{Flavor Oscillations}

Neutrino oscillations are now firmly established by measurements of
solar and atmospheric neutrinos and the KamLAND and K2K long-baseline
experiments\cite{Gonzalez-Garcia:2002dz,Fogli:2002au%
,Bahcall:2002ij,deHolanda:2002iv}.  Evidently the weak interaction
eigenstates $\nu_e$, $\nu_\mu$ and~$\nu_\tau$ are non-trivial
superpositions of three mass eigenstates $\nu_1$, $\nu_2$ and~$\nu_3$,
\begin{equation}
\pmatrix{\nu_e\cr \nu_\mu \cr \nu_\tau\cr}
=U \pmatrix{\nu_1\cr \nu_2 \cr \nu_3\cr}\,.
\end{equation}
The leptonic mixing matrix can be written in the canonical form
\begin{equation}
U=\pmatrix{1&0&0\cr0&c_{23}&s_{23}\cr0&-s_{23}&c_{23}\cr}
\pmatrix{c_{13}&0&e^{i\delta}s_{13}\cr0&1&0\cr
-e^{-i\delta}s_{13}&0&c_{13}\cr}
\pmatrix{c_{12}&s_{12}&0\cr-s_{12}&c_{12}&0\cr0&0&1\cr}\,,
\end{equation}
where $c_{12}=\cos\Theta_{12}$ and $s_{12}=\sin\Theta_{12}$ with
$\Theta_{12}$ the 12-mixing angle, and so forth.  One peculiarity of
3-flavor mixing beyond the 2-flavor case is a non-trivial phase
$\delta$ that can lead to CP-violating effects, i.e.\ the 3-flavor
oscillation pattern of neutrinos can differ from that of
anti-neutrinos.

The atmospheric neutrino oscillations essentially decouple from the
solar ones and are controlled by the 23-mixing that may well be
maximal (45$^\circ$). The solar case is dominated by 12-mixing that is
large but not maximal.  The Chooz reactor experiment provides an upper
limit on the small 13-mixing.  From a global 3-flavor analysis of all
data one finds the 3$\sigma$ ranges for the mass differences and
mixing angles summarized in Table~\ref{tab:mixingparameters}.

\begin{table}[h]
\tbl{Neutrino mixing parameters from a global analysis of
all experiments\protect\cite{Gonzalez-Garcia:2002dz}
(3$\sigma$ ranges).\vspace*{1pt}}
{\footnotesize
\begin{tabular}{|c|c|c|}
\hline
{} & {} & {} \\[-1.5ex]
Combination&Mixing angle $\Theta$&$\Delta m^2$ [meV$^2$]\\[1ex]
\hline
{} & {} & {}\\[-1.5ex]
12 & 27$^\circ$--42$^\circ$&55--190\\[1ex]
23 & 32$^\circ$--60$^\circ$&1400--6000\\[1ex]
13 & ${}<14^\circ$ &$\approx \Delta m_{23}^2$\\[1ex]
\hline
\end{tabular}\label{tab:mixingparameters}}
\vspace*{-13pt}
\end{table}

The only evidence for flavor conversions that is inconsistent with
this picture comes from LSND, a short-baseline accelerator
experiment. If the excess $\bar\nu_e$ counts are interpreted in terms
of $\bar\nu_\mu$-$\bar\nu_e$-oscillations, the allowed mixing
parameters populate two islands within\cite{Church:2002tc} $\Delta
m^2=0.2$--7~eV$^2$ and $\sin^22\Theta=(0.3\hbox{--}5)\times10^{-2}$.
One possibility to accommodate this $\Delta m^2$ with the atmospheric
and solar values is a fourth sterile neutrino appearing as an
intermediate state to account for the LSND measurements, although this
scheme is now almost certainly ruled out\cite{Maltoni:2002ac}.
Another solution is the radical conjecture that the masses of
neutrinos differ from those of anti-neutrinos, implying a violation of
the CPT symmetry\cite{Barenboim:2002ah}.

Therefore, if LSND is confirmed by the ongoing MiniBooNE
project\cite{Bazarko:2002pn} the observed flavor conversions imply
something far more fundamental than neutrino mixing. Many workers in
neutrino physics take the attitude that this would be too good to be
true and thus are skeptical about LSND. Either way, MiniBooNE is
crucial to clarify this fundamental point.

Assuming MiniBooNE will refute LSND so that there is no new
revolution, the mass and mixing parameters given in
Table~\ref{tab:mixingparameters} still leave many questions open. Is
the 23-mixing truly maximal while the 13-mixing is not?  How large is
the small 13-mixing angle? Is there a CP-violating phase?  Moreover,
it is possible that two mass eigenstates separated by the small
``solar'' mass difference could form a doublet separated by the large
``atmospheric'' difference from a lower-lying single state (``inverted
hierarchy'').

These issues will be addressed by long-baseline experiments involving
reactor and accelerator neutrinos.  KamLAND and K2K in Japan are
already taking data, while the Fermilab to Soudan and CERN to Gran
Sasso projects, each with a baseline of 730~km, are under
construction. Future projects involving novel technologies
(superbeams, neutrino factories,
beta-beams)\cite{NuFactCERN,NuFactWorkingGroup} and their physics
potential\cite{Barger:2000nf,Cervera:2000kp,Freund:2001ui} are being
discussed.  The ``holy grail'' of these efforts is finding leptonic CP
violation.

\section{Neutrino Dark Matter and Cosmic Structure Formation}

The most direct limit on the overall mass scale $m_\nu$ derives from
tritium experiments searching for a deformation of the $\beta$
end-point spectrum. The final limit from Mainz and Troitsk
is\cite{Weinheimer:2002rs}
\begin{equation}\label{eq:tritium}
m_\nu<2.2~{\rm eV}\qquad\hbox{(95\% CL)}\,.
\end{equation}
This number is much larger than the mass splittings, obviating the
need for a detailed interpretation in terms of mixed neutrinos.  In
future, the KATRIN experiment$\,$\cite{Weinheimer:2002rs} is expected
to reach a sensitivity of 0.35~eV.

Traditionally it is cosmology that provides the most restrictive
$m_\nu$ limits. Standard big bang cosmology predicts a present-day
density of
\begin{equation}\label{eq:numberdensity}
n_{\nu\bar\nu}=\frac{3}{11}\,n_\gamma \approx 112~{\rm cm}^{-3}
\quad\hbox{per flavor.} 
\end{equation}
This translates into a cosmic neutrino mass fraction of
\begin{equation}\label{eq:Omega_nu}
\Omega_\nu h^2=\sum_{i=1}^3 \frac{m_i}{92.5~{\rm eV}}\,,
\end{equation}
where $h$ is the Hubble parameter in units of $100~{\rm
km~s^{-1}~Mpc^{-1}}$. The oscillation experiments imply $m_\nu>40$~meV
for the largest neutrino mass eigenstate so that
$\Omega_\nu>0.8\times10^{-3}$ if $h=0.72$.  On the other hand, the
tritium limit Eq.~(\ref{eq:tritium}) implies $\Omega_\nu<0.14$ so that
neutrinos could still contribute significantly to the dark matter.

This possibility is severely constrained by large-scale structure
observations. Neutrino free streaming in the early universe erases
small scale density fluctuations so that the hot dark matter fraction
is most effectively constrained by the small-scale power of the cosmic
matter density fluctuations.  The recent 2dF Galaxy Redshift Survey
data imply\cite{Hannestad:2003xv,Elgaroy:2003yh}
\begin{equation}\label{eq:openbias}
\sum m_\nu < 1.0~{\rm eV} 
\qquad\hbox{(95\% CL)}\,.
\end{equation}
To arrive at this limit other cosmological data were used, notably the
angular power spectrum of cosmic microwave background radiation as
measured by WMAP as well as reasonable priors on other parameters such
as the Hubble constant. If one includes the more problematic
Lyman-$\alpha$ forest data to constrain the small-scale power of the
matter density the limit improves to\cite{Spergel:2003cb} $\sum m_\nu
< 0.69~{\rm eV}$. The dependence of such limits on priors and other
assumptions is discussed in the cited
papers\cite{Hannestad:2003xv,Elgaroy:2003yh}.

In future the Sloan Digital Sky Survey will improve the galaxy
correlation function while additional CMBR data from WMAP and later
from Planck will improve the matter power spectrum, enhancing the
cosmological $m_\nu$ sensitivity\cite{Hu:1997mj,Hannestad:2002cn}.
Especially promising are future weak lensing
data\cite{Hu:2002rm,Abazajian:2002ck} that may come surprisingly close
to the lower limit $\sum m_\nu>40~{\rm meV}$ implied by the
atmospheric neutrino oscillations.

While the progress in precision cosmology has been impressive one
should keep worrying about systematic effects that do not show up in
statistical confidence levels. Even when the cosmological limits are
nominally superior to near-future experimental sensitivities, there
remains a paramount need for independent laboratory experiments.

\section{How Many Neutrinos in the Universe?}

To translate a laboratory $m_\nu$ measurement or limit into a hot dark
matter fraction $\Omega_\nu$ and the reverse one usually assumes the
standard cosmic neutrino density
Eq.~(\ref{eq:numberdensity}). However, thermal neutrinos in the early
universe are characterized by unknown chemical potentials $\mu_\nu$ or
degeneracy parameters $\xi_\nu=\mu_\nu/T$ for each flavor.  While the
small baryon-to-photon ratio $\sim 10^{-9}$ suggests that all
degeneracy parameters are small, large asymmetries between neutrinos
and anti-neutrinos could exist and vastly enhance the overall density.

The recent WMAP measurement of the CMBR angular power spectrum
provides new limits on the cosmic radiation
density\cite{Hannestad:2003xv,Crotty:2003th,Pierpaoli:2003kw}.
However, the most restrictive limits on neutrino degeneracy parameters
still obtain from big-bang nucleosynthesis (BBN) that is affected in
two ways.  First, a larger neutrino density increases the primordial
expansion rate, thereby increasing the neutron-to-proton freeze-out
ratio $n/p$ and thus the cosmic helium abundance.  Second, electron
neutrinos modify $n/p\propto\exp(-\xi_{\nu_e})$. Depending on the sign
of $\xi_{\nu_e}$ this effect can compensate for the expansion-rate
effect of ${\nu_\mu}$ or ${\nu_\tau}$ so that no significant BBN limit
on the overall neutrino density obtains\cite{Kang:xa}.  If
$\xi_{\nu_e}$ is the only chemical potential, the observed helium
abundance yields $-0.01<\xi_{\nu_e}<0.07$.

However, neutrino oscillations imply that the individual flavor lepton
numbers are not conserved so that in thermal equilibrium there is only
one chemical potential for all flavors.  If equilibrium is achieved
before $n/p$ freeze-out, the restrictive BBN limit on $\xi_{\nu_e}$
applies to all flavors, $|\xi_\nu|<0.07$, fixing the cosmic neutrino
density to within about 1\%.  The approach to flavor equilibrium by
neutrino oscillations and collisions was recently studied%
\cite{Lunardini:2000fy,Dolgov:2002ab,Wong:2002fa,Abazajian:2002qx}.
The details are subtle due to the large weak potential caused by the
neutrinos themselves, causing the intriguing phenomenon of
synchronized flavor
oscillations\cite{Samuel:1993uw,Pastor:2001iu,Pastor:2002we}.

The bottom line is that effective flavor equilibrium before $n/p$
freeze-out is reliably achieved only if the solar oscillation
parameters are in the favored LMA region.  Now that KamLAND has
confirmed LMA, for the first time BBN provides a reliable handle on
the cosmic neutrino density.  As a consequence, for the first time the
relation between $\Omega_\nu$ and $m_\nu$ is uniquely given by the
standard expression Eq.~(\ref{eq:Omega_nu}).

\section{Neutrino Majorana Masses and Leptogenesis}

The neutrino contribution to the dark matter density is
negligible. Intriguingly, however, they may play a crucial role for
the baryon asymmetry of the universe (BAU) and thus the presence of
ordinary matter\cite{Fukugita:1986hr}.  The main ingredients of this
leptogenesis scenario are those of the usual see-saw mechanism for
small neutrino masses.  The ordinary light neutrinos have right-handed
partners with large Majorana masses.  The left- and right-handed
states are coupled by Dirac mass terms that obtain from Yukawa
interactions with the Higgs field.  The heavy Majorana neutrinos will
be in thermal equilibrium in the early universe. When the temperature
falls below their mass, their equilibrium density becomes
exponentially suppressed. However, if at that time they are no longer
in thermal equilibrium, their abundance will exceed the equilibrium
distribution. The subsequent out-of-equilibrium decays can lead to the
net generation of lepton number.  CP-violation is possible by the
usual interference of tree-level with one-loop diagrams.  The
generated lepton number excess will be re-processed by standard-model
sphaleron effects which respect $B-L$ but violate $B+L$. It is
straightforward to generate the observed BAU by this mechanism.

The requirement that the heavy Majorana neutrinos freeze out before
they get Boltzmann suppressed implies an upper limit on the same
parameter combination of Yukawa couplings and heavy Majorana masses
that determines the observed small neutrino
masses\cite{Buchmuller:2000as}.  Most recently, a robust upper limit
on all neutrino masses of
\begin{equation}
m_\nu<120~{\rm meV}
\end{equation}
was claimed\cite{Buchmuller:2003gz}.  Degenerate neutrinos with a
``large'' common mass scale of, e.g., 400~meV require a very precise
degeneracy of the heavy Majorana masses to better than $10^{-3}$.

A necessary ingredient for this mechanism is the Majorana nature of
neutrino masses that can be tested in the laboratory by searching for
neutrinoless $2\beta$ decay. This process is sensitive to
\begin{equation}
\langle m_{ee}\rangle=\left| 
\sum_{i=1}^3\lambda_i \left|U_{ei}\right|^2 m_i\right|
\end{equation}
with $\lambda_i$ a Majorana CP phase. Therefore, we have two
additional physically relevant phases beyond the Dirac phase $\delta$
of the previously discussed mixing matrix. If neutrinos have Majorana
masses their mixing involves three mass eigenstates, three mixing
angles, and three physical phases.

Actually, several members of the Heidelberg-Moscow collaboration have
claimed first evidence for this
process\cite{Klapdor-Kleingrothaus:2001ke,Klapdor-Kleingrothaus:pr},
implying a 95\% CL range of $\langle m_{ee}\rangle=110$--560~meV.
Uncertainties of the nuclear matrix element can widen this range by up
to a factor of~2 in either direction.  The significance of this
discovery has been fiercely critiqued by many experimentalists working
on other $2\beta$ projects\cite{Aalseth:2002dt}.  Even when taking the
claimed evidence at face value the statistical significance is only
about 97\%, too weak for definitive conclusions. More sensitive
experiments are needed and developed to explore this range of Majorana
masses\cite{Cremonesi:2002is}.

\section{High-Energy Neutrinos From Astrophysical Sources}

The 2002 Physics Nobel Prize was awarded, in part, to Raymond Davis
and Masatoshi Koshiba ``for pioneering contributions to astrophysics,
in particular for the detection of cosmic neutrinos.''  Unfortunately,
the observed sources remain limited to the Sun and Supernova 1987A,
apart from the diffuse flux of atmospheric neutrinos. This situation
could radically change in the near future if the high-energy neutrino
telescopes that are currently being developed begin to discover
astrophysical point sources.

The spectrum of cosmic rays reaches to energies of at least
$3\times10^{20}~{\rm eV}$, proving the existence of cosmic sources for
particles with enormous energies\cite{Sigl:ih,Halzen:2002pg}.  Most of
the cosmic rays appear to be protons or nuclei so that there must be
hadronic accelerators both within our galaxy and beyond.  Wherever
high-energy hadrons interact with matter or radiation, the decay of
secondary pions produces a large flux of neutrinos At the source one
expects a flavor composition of $\nu_e:\nu_\mu:\nu_\tau=1:2:0$, but
the observed oscillations imply equal fluxes of all flavors at Earth.
High-energy neutrino astronomy offers a unique opportunity to detect
the enigmatic sources of high-energy cosmic rays because neutrinos are
neither absorbed by the cosmic photon backgrounds nor deflected by
magnetic fields.

While there are many different models for possible neutrino
sources\cite{Halzen:2002pg,Waxman:2002wp}, the required size for a
detector is generically 1~km$^3$. The largest existing neutrino
telescope, the AMANDA ice Cherenkov detector at the South Pole, is
about 1/10 this size. It has not yet observed a point source, but the
detection of atmospheric neutrinos shows that this approach to
measuring high-energy neutrinos works well\cite{Ahrens:2002hh}. It is
expected that this instrument is upgraded to the full 1~km$^3$ size
within the next few years under the name of
IceCube\cite{Halzen:2003yh}.  Similar instruments are being developed
in the Mediterranean\cite{Carr:da}.  Moreover, air-shower arrays for
ordinary cosmic rays may detect very high-energy neutrinos by virtue
of horizontal air showers\cite{Letessier-Selvon:2002fh}.  Although
this field is in its infancy, it holds the promise of exciting
astrophysical discoveries in the foreseeable future.

\section{Supernova Neutrinos}

The observation of neutrinos from the supernova (SN) 1987A in the
Large Magellanic Cloud was a milestone for neutrino astronomy, but the
total of about 20 events in the Kamiokande and IMB detectors was
frustratingly sparse. The chances of observing a galactic SN are small
because SNe are thought to occur with a rate of at most a few per
century. On the other hand, many neutrino detectors and especially
Super-Kamiokande have a rich physics program for perhaps decades to
come, notably in the area of long-baseline oscillation experiments and
proton decay searches.  Likewise, the south pole detectors may be
active for many decades and would be powerful SN
observatories\cite{Ahrens:2001tz}. Therefore, it remains worthwhile to
study what can be learned from a high-statistics SN observation.

The explosion mechanism for core-collapse SNe remains unsettled as
long as numerical simulations fail to reproduce robust explosions. A
high-statistics neutrino observation is probably the only chance to
watch the collapse and explosion dynamics directly and would allow one
to verify the standard delayed explosion scenario\cite{Totani:1997vj}.
The neutrinos arrive a few hours before the optical explosion so that
a neutrino observation can serve to alert the astronomical community,
a task pursued by the Supernova Early Warning System
(SNEWS)\cite{SNEWS}.  For particle physics, many of the limits based
on the SN~1987A neutrino signal\cite{Raffelt:1999tx} would improve and
achieve a firm experimental and statistical basis.  On the other hand,
the time-of-flight sensitivity to the neutrino mass is in the range of
a few eV,\cite{Beacom:2000qy} not good enough as the ``$m_\nu$
frontier'' has moved to the sub-eV range.

Can we learn something useful about neutrino mixing from a galactic SN
observation? This issue has been addressed in many recent studies%
\cite{Dighe:1999bi,Lunardini:2003eh,Takahashi:2002cm,Takahashi:2001dc},
and the answer is probably yes, depending on the detectors operating
at the time of the SN, their geographic location, and the neutrino
mixing scenario, i.e.\ the magnitude of the small mixing angle
$\Theta_{13}$ and the ordering of the masses. Any observable
oscillation effects depend on the spectral and flux differences
between the different flavors.  We have recently shown that previous
studies overestimated these
differences\cite{Raffelt:ai,Buras:2002wt,Keil:2002in} because
traditional numerical simulations used a schematic treatment of
$\nu_\mu$ and $\nu_\tau$ transport.  Distinguishing, say, between the
normal and inverted mass ordering remains a daunting task at
long-baseline experiments. Therefore, a future galactic SN observation
may still offer a unique opportunity to settling this question.

The relic flux from all past SNe in the universe is observable because
it exceeds the atmospheric neutrino flux for energies below
30--40~MeV. Recently Super-Kamiokande has reported a limit that
already caps the more optimistic
predictions\cite{Malek:2002ns}. Significant progress depends on
suppressing the background caused by the decay of sub-Cherenkov muons
from low-energy atmospheric neutrinos.  One possibility is to include
an efficient neutron absorber such as gadolinium in the detector that
would tag the reactions $\bar\nu_e+p\to n+e^+$.\cite{Vagins2003} If
this approach works in practice the detection of relic SN neutrinos
has come within experimental reach.

\section{Conclusions}

After neutrino oscillations have been established, the next challenge
is to pin down the as yet undetermined elements of the mixing matrix
and the absolute masses and mass ordering. Long-baseline experiments
can address many of these questions and may even discover leptonic
CP-violation. The Majorana nature of neutrinos can be established in
$2\beta$ experiments if the $0\nu$ decay mode can be convincingly
observed.  Majorana neutrinos with masses ${}<120$~meV fit nicely into
the leptogenesis scenario for creating the baryon asymmetry of the
universe so that neutrinos may well be responsible for the ordinary
matter in the universe. Their contribution to the dark matter is
non-zero but negligible. Still, precision observations of cosmological
large-scale structure remain the most powerful tool to constrain the
absolute mass scale, although independent laboratory confirmation
remains crucial.  In the past, neutrino oscillation physics was
dominated by solar and atmospheric neutrinos, but long-baseline
experiments are about to ``take over.''  In future, neutrinos from
natural sources are likely more important as astrophysical messengers
while oscillation physics will mostly be done in the laboratory.  The
search for astrophysical point sources with high-energy neutrino
telescopes may soon open a new window to the universe. Observing a
high-statistics neutrino signal from a future galactic supernova
remains perhaps the most cherished prize for low-energy neutrino
observatories.  Meanwhile the search for the cosmic relic neutrinos
from all past supernovae has become a realistic possibility. Neutrino
physics and astrophysics will remain fascinating for a long time to
come!

\section*{Acknowledgments}

This work was supported, in part, by the Deutsche
Forschungsgemeinschaft under grant No.\ SFB-375 and by the European
Science Foundation (ESF) under the Network Grant No.~86 Neutrino
Astrophysics.


\end{document}